# The effects of urbanization on the temperature over the Beijing-Tianjin-Hebei region under changing climate

Xinyi Ai[1], Yichuan Huang[2], Qizhen Dong[3], Kai Yang[4], Danhong Dong[4], Xichen Li[4]

1. The High School Affiliated to Renmin University of China

2. Capital Normal University High School

3. Hangzhou Foreign Languages School

4. State Key Laboratory of Numerical Modeling for Atmospheric Sciences and Geophysical Fluid Dynamics, Institute of Atmospheric Physics, Chinese Academy of Sciences, Beijing

Corresponding author address: Dr. Danhong Dong, LASG/IAP/CAS, P. O. Box 9804, Beijing 100029, China. Email: dongdanhong@mail.iap.ac.cn

# Abstract

Urbanization has a large impact on human society. Besides, global warming has become an increasingly popular topic since global warming will not only play a significant role in human beings' daily life but also have a huge impact on the global climate. Under the background of global warming, the impacts of urbanization may have differences, which is the topic this project mainly discusses. In this study, the monthly air temperature data from 2 meteorological stations in Beijing under the different underlying surface: urban and suburban for a period of 1960-2014 were analyzed. Besides, two years, 1993 and 2010, selected with the different circulation conditions have been simulated by weather forecasting models. We conducted experiments using the Weather Research and Forecasting (WRF) model to investigate the impacts of urbanization under different circulation conditions on the summer temperature over the Beijing-Tianjin-Hebei Region (BTHR). The results indicate that the effect of urbanization under low-pressure system is greater than that under high-pressure system. That's because the difference of total cloud cover between the urban area and the suburban area is greater under low-pressure system in 1993. This research will not only increase our understanding of the effect of urbanization but also provide a scientific basis to enhance the ability to prevent disasters and reduce damages under the global warming background.

# 1. Introduction

China is the largest developing country in the world, and its economy has undergone a continuous expansion ever since 1978 (Yin et al. 2014). In recent decades, more and more rural residents choose to move to big cities for better job opportunities with a higher salary, medical care, education recourses, and living conditions. Grimm et al. (2008) indicated that the percentage of urban citizens in China has already exceeded 50% by now and will reach 60% by the end of the year 2030. BTHR, as the most essential economics and political center of China with a concentrated population, experienced rapid urbanization significantly. Wu et al. (2015) mentioned that from 1980 to 2010, the average annual urban growth rates of Beijing, Tianjin and Hebei are 3.7%, 4.7%, and 3.2% respectively, and will be tripled by the end of 2030. Such rapid urbanization, with increasing build-up areas, induces artificial changes in land use to modify surface and boundary layer atmospheric properties. Besides, due to the urbanization, the increase of air pollution and vehicles will increase the anthropogenic heat and aerosol, accompanied by the changes of climate and hydrological cycle (Grimm et al. 2008; Lubchenco et al. 1997).

In general, urbanization affects the temperature and atmospheric circulation through two types: land cover changes and the increase of anthropogenic pollutant emissions (Zhong et al. 2015; Zhong et al. 2017). The urban heat island (UHI) effect is the phenomenon that the surface temperature in the urban area is higher than the suburban area. It is a popular topic of anthropogenic modification on Earth system to explain the effect of urbanization on temperature and regional climate (Akbari and Kolokotsa 2016; Arnfield 2003; Gong et al. 2012; IPCC 2007; Kaloustian and Diab 2015; Mirzaei 2015; Oke 1973; O'Loughlin et al. 2012; Patz et al. 2005; VoogtandOke 2003; Zhou et al. 2014). UHI is caused by the construction of tall buildings, the lack of vegetation, the trapped heat and so on. Furthermore, since the amount of solar radiation energy obtained by the surface is determined by the albedo, the transformation of the underlying surface is another factor that causes UHI as well. The average annual temperature of the BTHR rises dramatically because of urbanization, aggravates UHI, and therefore strongly influences the climate and the national economy. Besides, the urbanization also has effects on climate (Arnfield 2003; Dixon and Mote 2003; IPCC 2007; Jin et al. 2005a; Jin et al. 2005b; Shepherd 2005), which would bring more serious problems when adding global warming into consideration (IPCC 2007; McCarthy et al. 2010; Patz et al. 2005). The change of underlying surface type has an important influence on local temperature, but the process may be different under different atmospheric circulation backgrounds.

Using the regional model, this project not only studies the influence of urban underlying surface types on air temperature in the BTHR, but also further focuses on the influence of different circulation background underlying surface types on local air temperature, and discusses the physical process involved. This will not only improve our understanding of the impact of urbanization on climate, but also provide a scientific basis for China to prevent the negative consequences include human health and well-beings like an increase in morbidity, mortality, and risk of violence (Gong et al. 2012; O'Loughlin et al. 2012).

This paper is organized as followed. Section 2 describes the datasets and methods used in this paper. In Section 3, we show the main results of our study. The conclusions are made and represented in Section 4.

**2. Data and methods**

Monthly 2 m observational data of air temperature over Beijing during the period of 1960-2014 from the China Meteorological Administration (CMA) via http://data.cma.cn/ was used to show the difference of temperature between urban area and suburban area. The Hadley Centre-Climate Research Unit merged land air temperature and ocean surface temperature (MLOST) version 4.4.0.0 median (HadCRUT4) data (Morice et al. 2012) from 1960-2014 are used to calculate the global mean air temperature. The geopotential height data from the European Centre for Medium-Range Weather Forecasts (ECMWF) Interim reanalysis (ERA-Interim) dataset (Dee et al. 2011) has been used to calculate the regional mean 500 hPa geopotential height over the BTHR, which can represent the circulation condition of mid-troposphere. Digital elevation model (DEM) and land use cover over Beijing are provided by the global land cover with a spatial resolution of 30 m (more details are provided at www.gscloud.cn and www.globallandcover.com).

To investigate the effects of the changing of land use type on air temperature over the BTHR, we use the Weather Research and Forecast model (WRF) version 3.9.0 (Skamarock et al. 2008) to simulate the climate over the BTHR, and emphasize the influences of land use types through several sensitivity experiments. WRF is a flexible atmospheric simulation system and has been widely used in numerical weather prediction and climate research over East Asia (Zhang et al. 2011). It has multiple options for atmospheric physical processes, which include the Planetary boundary layer (PBL), longwave and shortwave radiation, microphysics, cumulus parameterization, and land surface physics. In this study, we use the National Center for Atmospheric Research Community Atmosphere Model shortwave and longwave schemes (Collins et al. 2006), the WRF-Single-Moment three-class scheme for the microphysics parameterization (Hong and Lim 2006), the Kain-Fritsch convective parameterization scheme as the cumulus convective scheme (Kain 2004), and the Yonsei University scheme in the PBL (Hong and Lim 2006), which are mainly based on earlier simulations of the regional climate over East Asia (Gao et al. 2017; Li et al. 2018; Zhang et al. 2011). ERA-interim data with a spatial resolution of 1°×1° and a time resolution of 6 hours are used to provide the lateral boundary conditions and sea surface temperature data for the WRF simulation.

The WRF simulations were set with a 5-km horizontal resolution. Two years with different circulation condition, which was represented by the mean values of 500 hPa geopotential height over the BTHR in summer, are selected for sensitivity experiments. Each year has two runs from May 1st to August 31th. The simulations from May 1st to May 31st is for "spin up", only the data from June 1st to August 31st (summer) are used for investigation. For the two sensitivity experiments of each year, everything is the same except that the land use type of one run is "Urban and Built-Up" and that of the other run is " Mixed Forests".

## 3. Results

We use the observations of the surface air temperature data at the urban station and the suburban station of Beijing, which is one of the most important cities and the most urbanized cities in the BTHR, as an example to show the differences of the surface air temperature over the urban regions and suburban regions. The spatial distributions of the altitudes and land use types over Beijing are shown in Fig. 1. Beijing is located at the northwest edge of the North China plain, backed by the Taihang mountains and Yanshan mountains. With its southeast about 150 kilometers away from the Bohai Sea, Beijing is facing the vast north China plain. Beijing's terrain is generally higher in the northwest part and lower in the southeast part. Two meteorological stations with similar altitude are selected to limit the effect of altitude on temperature (Fig.1a). These two stations are located in different areas. One is in the urban area and the other one is in grassland area (Fig. 1b).

Figure 2 shows the time series of the summer mean air temperature anomalies in two observational stations and the mean value of the global averaged summer air temperature anomalies for the period of 1960-2014. We can find that the global mean temperature anomalies generally show an increasing trend, which shifts from -0.2 to 0.8 and increases relatively constantly. While the mean temperature anomalies in the urban area of Beijing shows higher interannual variation. At the beginning of the 21st century, the temperature anomalies in urban is significantly higher than the global mean level (Fig. 2a). We also find that the urban mean temperature anomalies and the suburban mean temperature anomalies generally show an increasing trend, and the interannual variation of both the urban and the suburban temperature anomalies are higher than the global mean level. In particular, in the early 21st century, the urban temperature anomalies is significantly higher than the suburban temperature anomalies, which shows the effects of urbanization (Fig. 2b).

To investigate the effects of urbanization on the temperature under changing circulation conditions, we calculate the regional mean of the 500 hPa geopotential height over the BTHR (Fig. 3). Two years (1993 and 2010) have been selected as the years with high-and low-pressure systems, respectively.

Figure 4 shows the summer mean geopotential heights of our WRF simulations in 1993 and 2010 over the BTHR. Comparing the geopotential heights in 1993 to that of 2010 with two different land use types, it is obvious that the BTHR is controlled by higher atmospheric pressure in 2010, which is consistent with observation. The geopotential heights of 2010 are about 60 meters higher than the geopotential heights in 1993. The geopotential heights show less difference under different land use types, which means that the changing of the land use barely has effects on the atmospheric pressure.

Figure 5 shows a significant disparity of the surface air temperature over the BTHR under different land use types. As we assumed, the summer mean temperature of the urban areas is considerably higher than those of the suburban areas. This can be due to the lower albedo of city landscapes and the UHI effect. Generally, the summer mean temperature over the BTHR is about 1.5℃ to 2.0℃ higher in 2010 than 1993, which means the summer mean temperature is positively influenced by the atmospheric pressure (Fig. 5). When a region is controlled by a high atmospheric pressure, the air tends to flow downward to the ground, thus the water vapor can hardly deposit to form clouds. The downward shortwave radiation is stronger without clouds, which is why the temperature is normally high under high-pressure system. Comparing the differences between the two cases, we find that the differences of temperature between urban areas and suburban areas are smaller under high-pressure environment than that under low-pressure environment (Fig. 5e and 5f). We then try to explore the reasons for this phenomenon.

As we have discovered, the albedo is much higher in the suburban areas than in the urban areas. Comparing the albedo in 1993 to the albedo in 2010 for two cases, the urban area did not present a considerable change as the atmospheric pressure increases (Fig. 6a-6d). However, the albedo in suburban area shows an increase in a wide area, about 1.125%. So, we inferred that the albedo of suburban areas is more likely to change when atmospheric pressure is changing. This can be a result of the fact that the plants, which have great effects on albedo, can be influenced by the temperature. For example, the plants will close their pores in hot conditions to decrease the amount of water evaporation.

Comparing the cloud cover over urban areas to those over suburban areas, it is apparent that the cloud cover over suburban areas is more than that in urban areas (Fig.

6e-6h). When comparing the cloud cover in 1993 to the cloud cover in 2010, we find that the differences of the cloud cover for two cases in 1993 is significantly larger than that in 2010, which means that the cloud cover is much more sensitive to land use type under low-pressure system. The formation of cloud needs water vapors and lifting movement. When a region is controlled by high-pressure system, the downward movement will limit the formation of cloud, even the land surface has enough water vapor. So, the cloud and albedo are more sensitive to land use type under low-pressure system, which leads to the downward shortwave radiation and air temperature are more sensitive to land use type under low-pressure system.

Fig. 7 shows that the difference of downward shortwave radiation between urban area and suburban area in 1993 is larger than that in 2010, which is because the difference between the total cloud cover in the urban area and the suburban area is greater in 1993 than that in 2010 (Fig. 6). As the quantity of clouds increases, the amount of short wave that the clouds reflect increases as well and vice versa.

The general process of energy balance and the effects of urbanization are shown in Fig. 8. The suburban area is more conducive to form clouds than the urban area dues to the fact that it's covered with grasses, shrubs, and trees, which provide lots of water vapors by transpiration and could later reserve them by various ways. Thus, fewer clouds are formed in the urban area than in the suburban area. Therefore, fewer solar radiation is reflected by the clouds in the urban area than in the suburban area. On account of this reason, there's more solar radiation that reaches the urban surface, and as discussed above, the albedo of the suburban surface is greater than that of the urban surface, more solar radiation is absorbed by the urban surface than the suburban surface. The temperature in the urban area is consequently higher compared to that in the suburban area due to this reason and hence the amount of surface radiation emitted by urban surface is greater than that emitted by suburban surface. Owing to the reason that there's obviously more cars and air conditioners in urban area, carbon dioxide and ozone—two of the major greenhouse gases that contribute significantly to the greenhouse effect—cumulate in the atmosphere above the urban area and form a protective shield together with other gases and aerosols around the earth, which absorbs the surface radiation and then emits more back radiation to the ground and less longwave radiation to the atmosphere in the urban area than in the suburban area and again leads to the rise in temperature in the urban areas. This is a complete cycle that explains how urbanization affects the overall temperature in the urban area.

**4. Conclusions and discussions**

In this study, we investigate the effect of urbanization over the BTHR on air temperature under different circulation background by doing two sensitivity experiments using the WRF model. The year 1993 under the lowest pressure system and year 2010 under the highest-pressure system are selected for the experiments. We find that the influence of urbanization on temperature differs significantly under different circulation conditions, indicating that the urbanization effect on temperature cannot be generalized even in the same region. The main results are summarized as follows.

1) The temperature of 2010 is apparently greater than that of 1993 due to the reason that descending motion of the atmosphere increases under high-pressure system, which is not conducive to form clouds, and further causes more solar radiation to be absorbed by the ground.

2) 1993 and 2010 have a diverse value of differences in ground's albedo, and this difference is mainly caused by the case of the suburban area. That's because the change in temperature affects plants' rates of respiration and transpiration, which further lead to the distinct value of the ground's albedo.

3) The effect of urbanization under low-pressure system is greater than that under high-pressure system. That's because the difference of total cloud cover between the urban area and the suburban area is greater under low-pressure system in 1993. Under high-pressure system, there are more downdraft which limits the cloud formation even though there's more water vapor and droplets accumulated in the suburban area. Thus, more clouds are formed under low-pressure system, which results in a greater difference between total cloud cover between the urban area and the suburban area.

The results in our study suggest the potential sensitivity of temperature to urbanization under different circulation backgrounds to some extent, which may improve our understanding of the effects of urbanization on global climate in a warming world. However, the conclusions require further examinations with more case studies.

**Acknowledgements**

The ERA-Interim data used in this study were obtained from the ECMWF data server: http://apps.ecmwf.int/datasets/. The authors would like to thank the Institute of Atmospheric Physics, Chinese Academy of Sciences for providing data and facilities for this research.

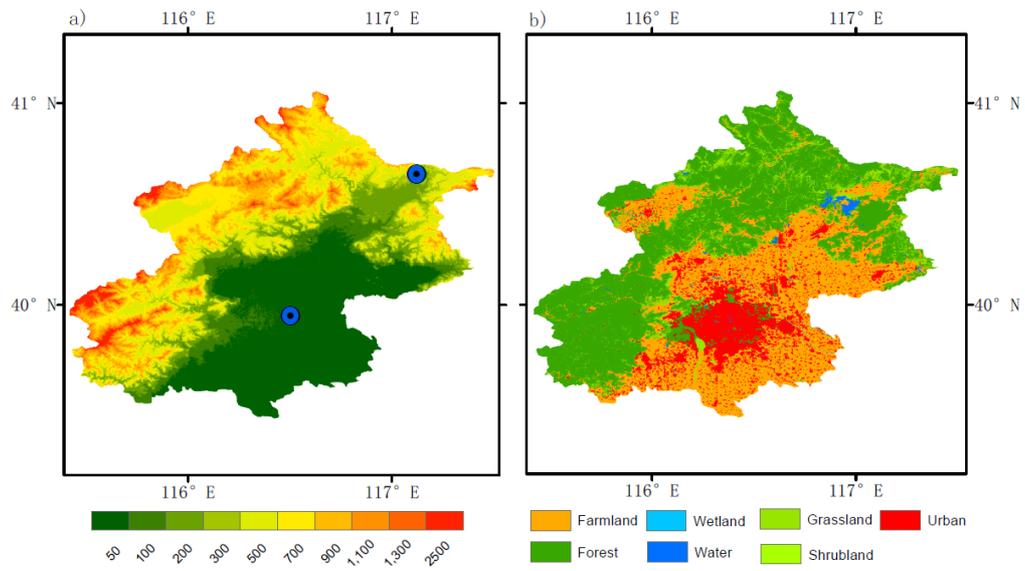

Fig. 1 a) Elevation (m) of Beijing and its distribution of meteorological stations (Two points refer to the urban station and the suburban station, respectively). b) Land use types of Beijing.

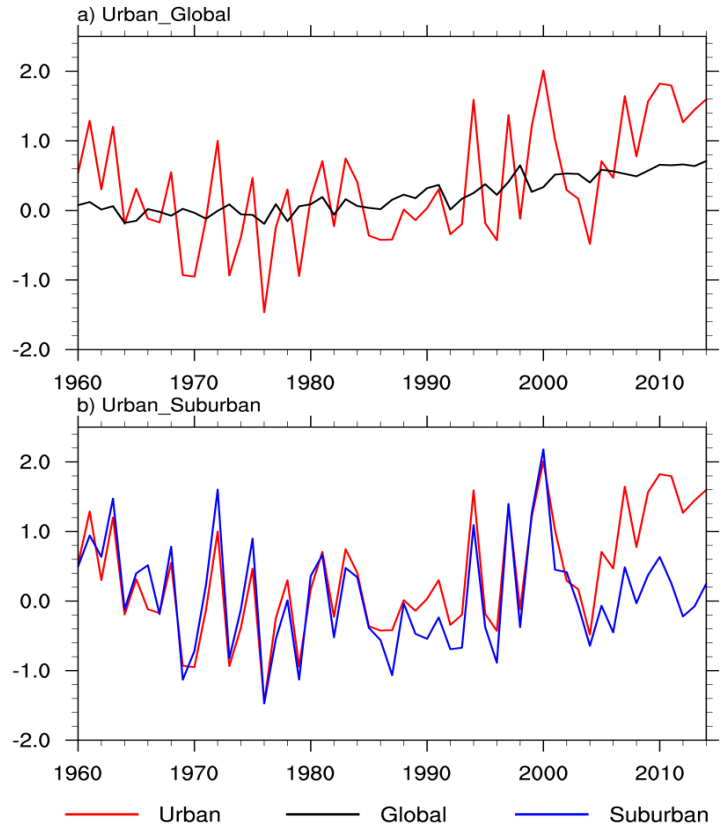

Fig. 2 a) Anomalies of summer mean temperature (k) of the Beijing urban station (red line) and the whole global (black line). b) Anomalies of summer mean temperature of the urban station (red line) and the suburban station (blue line) of Beijing. The anomalies are calculated comparing to the average values for the period of 1960 to 1990. The period of 1960-1990 has always been used as a time period that has been less influenced by human beings.

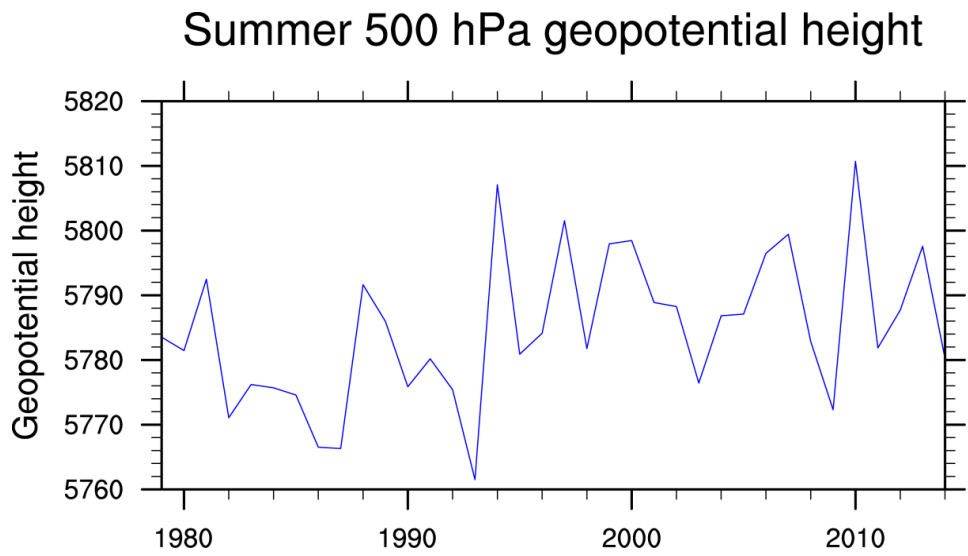

Fig. 3 Regional averaged 500 hPa geopotential height (m) field over the BTHR from 1979 to 2014.

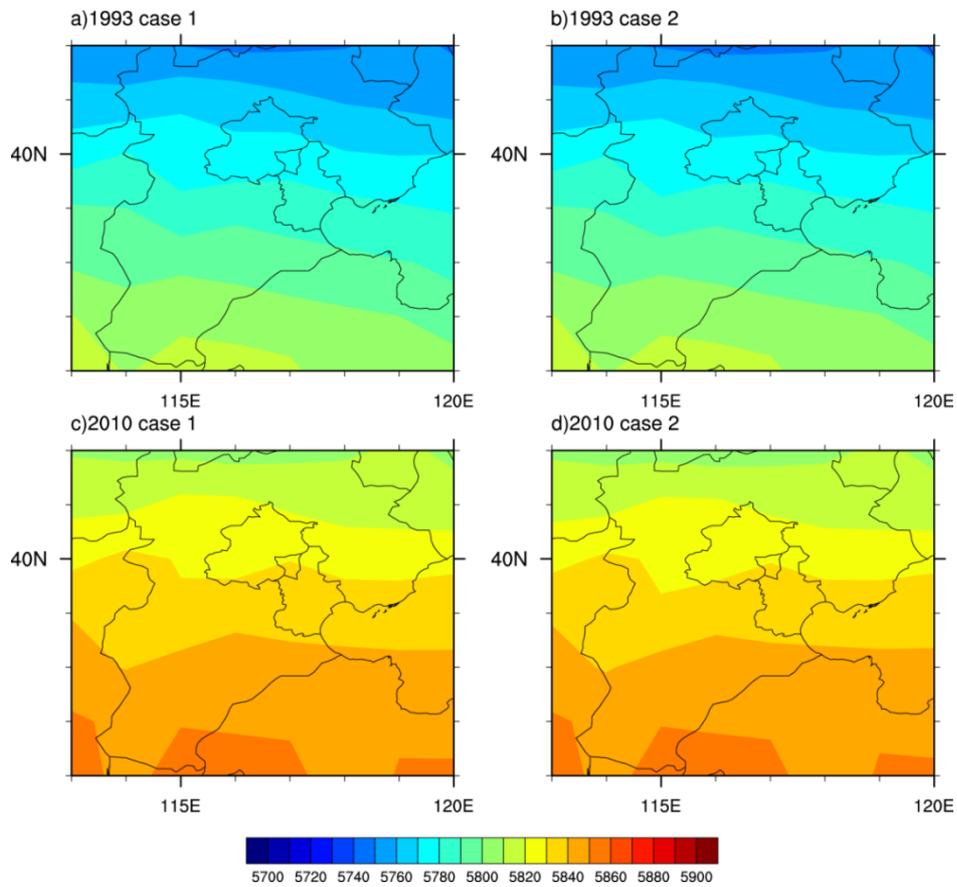

Fig. 4 Summer mean geopotential height (m) in 1993 over the BTHR of the a) urban case and b) suburban case; Summer mean geopotential height in 2010 over the BTHR of the c) urban case and d) suburban case.

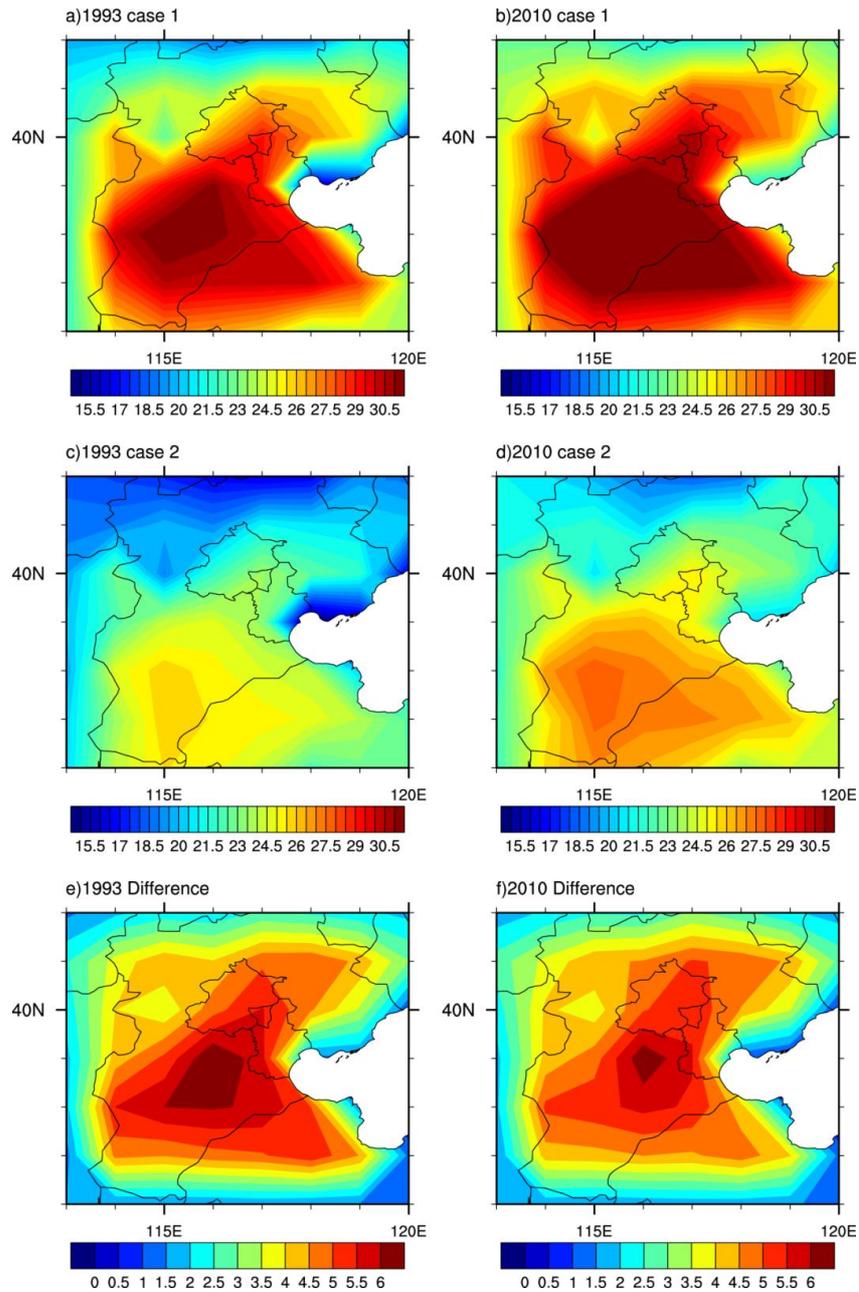

Fig. 5 Summer mean temperature in 1993 over the BTHR of the a) urban case and c) suburban case; Summer mean temperature in 2010 over the BTHR of the b) urban case and d) suburban case; The difference value between urban and suburban summer mean temperature over the BTHR in e) 1993 and f) 2010.

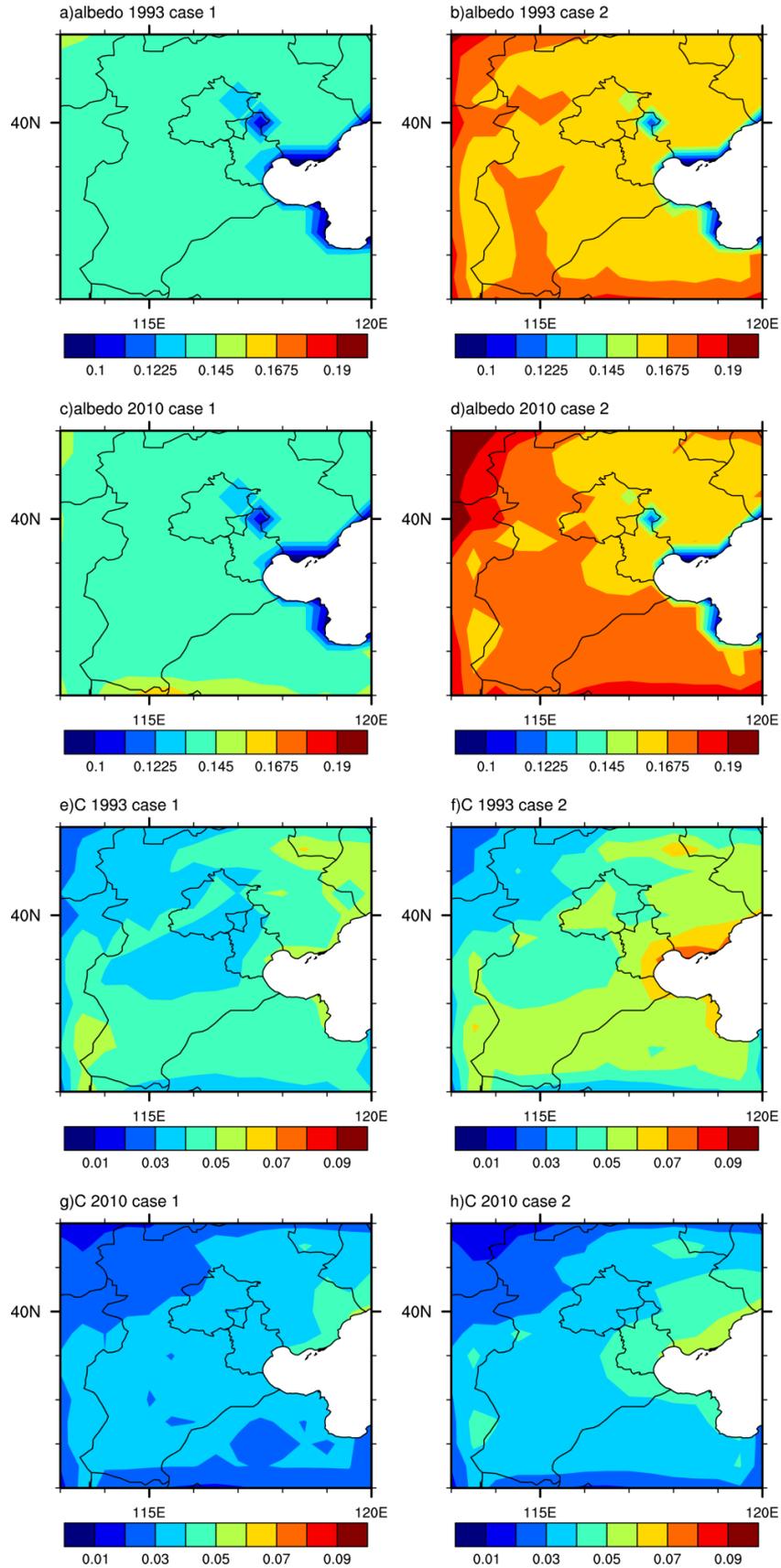

Fig. 6 Summer mean albedo in 1993 over the BTHR of the a) urban case and b)suburban

case; Summer mean albedo in 2010 over the BTHR of the c) urban case and d)suburban case; The amount of cloud formation in summer in 1993 over the BTHR of the e) urban case and f) suburban case; The amount of cloud formation in summer in 2010 over the BTHR of the g) urban case and h)suburban case.

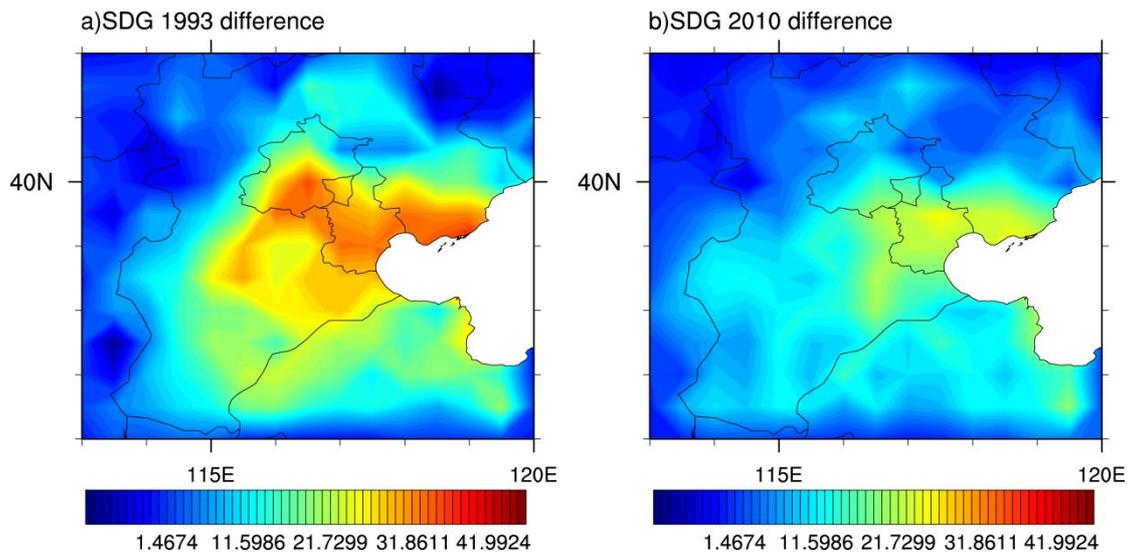

Fig. 7 The differences between downward short wave at the urban ground and suburban ground in a) 1993 and b) 2010.

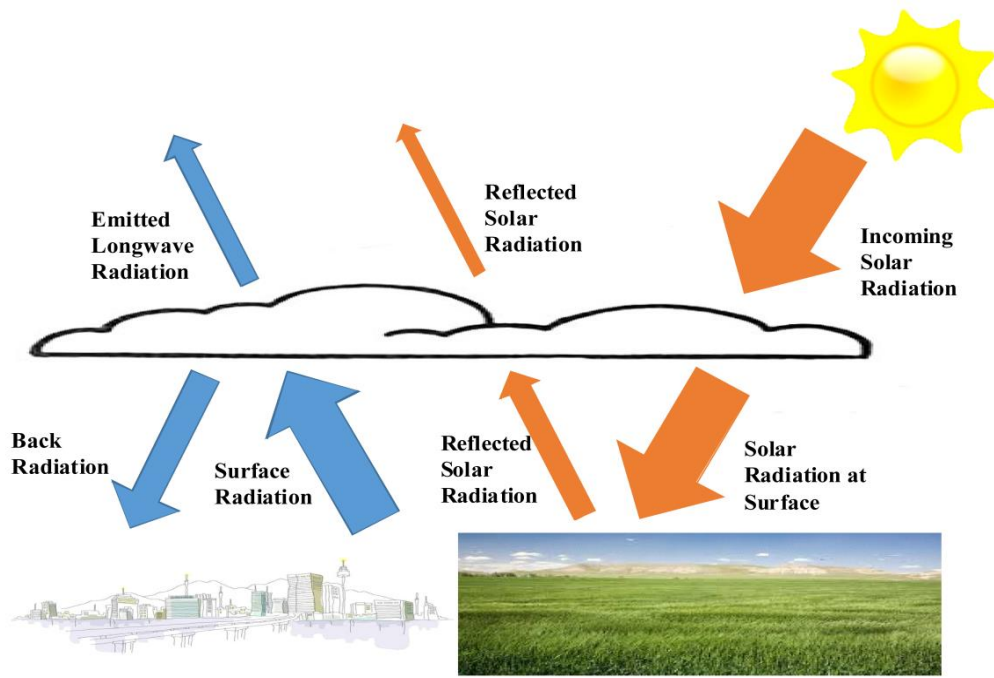

Fig. 8 A diagram that demonstrates the effect of urbanization